\documentclass[useAMS,usenatbib,usegraphicx]{mn2e}

\usepackage{color}

\title[Fe~K$\alpha$ Line in Hard X-ray Emitting Symbiotic Stars]{Fe~K$\alpha$ Line in Hard X-ray Emitting Symbiotic Stars}
\author[R.N.C. Eze]{R.N.C. Eze$^{1,2}$\thanks{E-mail:
romanus.eze@gmail.com; romanus.eze@unn.edu.ng (RNCE)} \\
$^{1}$Institute of Space and Astronautical Science,
Japan Aerospace Exploration Agency
3-1-1 Yoshinodai, Chou-ku, Sagamihara, \\ Kanagawa  252-0222, Japan \\
$^{2}$Department of Physics and Astronomy
University of Nigeria, Nsukka, 
Enugu State, Nigeria.}

\begin{document}

\date{Accepted 2013 October 08. Received 2013 October 08; in original form 2012 November 05}

\pagerange{\pageref{firstpage}--\pageref{lastpage}} \pubyear{2013}

\maketitle

\label{firstpage}

\begin{abstract}
The 6.4 keV iron emission line is typically created by irradiation of the neutral (or low ionized) iron by a hard X-ray source.  Whereas the 6.7 and 7.0 keV emission lines are mainly produced by photoionization and  collisional excitation in hot plasma, the 6.4 keV fluorescence line is typically a signature of either reflection from an accretion disk or absorption.  We have surveyed the emission using a collection of Suzaku observations of hard X-ray emitting symbiotic stars to better understand the geometry of these systems.  We find that they do not seem to have a single geometry, and that while absorption-induced fluorescence leads to some emission in three of the hard X-ray emitting symbiotic stars (hSSs) in our study, CH~Cyg, T~CrB and RT~Cru there are strong hints that significant 6.4 keV emission arises in the accretion disk irradiated by the hard X-rays from the boundary layer between the accretion disk and hot white dwarf in one of our sources, SS73~17. The 6.7 and 7.0 keV lines, however, are largely produced by collisional excitation in the vicinity of the compact white dwarf.

\end{abstract}



\section{Introduction}

Symbiotic stars are interacting binaries formed by a red giant star and a hot degenerate companion which accretes mass from the stellar wind of the red giant, forming a nebula surrounding the system which is typically detected via various optical emission lines. Detailed properties of hard X-ray emitting symbiotic stars are given by \citet{b9}, and references therein.

The 6.4 keV iron emission line is typically created by irradiation of the neutral (or low ionized) material (iron) by a hard X-ray source. This  line was first discovered in an astronomical context by \citet{b22} in OSO-8 X-ray observations of Cen A (NGC 5128). NGC 5128 is one of the closest radio galaxies and one of the first extragalactic objects to be identified as an X-ray source. X-ray binaries also showed K-shell transitions from cold neutral metals (e.g. \citet{b2}). The Ginga satellite also detected a strong Fe~K$\alpha$  fluorescence line in a number of Seyfert 1 galaxies (see e.g. \citet{b14,b21,b27,b29}). The Fe~K$\alpha$ line of active galactic nuclei, as well as different geometries and astrophysical conditions necessary for its emergence, was intensively studied in the 1990s (see e.g. \citet{b18,b11}).

The key results of these studies was that the fluorescent Fe~K$\alpha$ at 6.4 keV originates from the innermost part of accretion disk, close to the central compact object. The general consensus is that X-ray irradiation of the surface layers of the accretion disk in radio galaxies and AGN, especially Seyfert  galaxies gives rise to fluorescent Fe~K$\alpha$ emission line. As an extreme example, \citet{b1}  discovered a broad Fe~K$\alpha$ line emitted from a region which had inner radius close to the innermost stable circular orbit (R$_{ms}$) for a maximally spinning black hole with outer radius within 10 R$_{g}$ (where R$_{g}$ is the gravitational radius) using XMM-Newton observation of the broad- line radio galaxy 4C 74.26. The Fe~K$\alpha$  fluorescence line is typically strong, thanks to iron's high abundance and the large radiative transition rate for the 2p--1s transition. It can be found in the spectra of all types of accreting sources: binary black hole and neutron star systems, cataclysmic variable stars, hard X-ray emitting symbiotic stars and AGN. The ASCA satellite with its excellent spectral resolution and sensitivity, was used to obtain the first convincing proof for the existence of a relativistically broadened Fe~K$\alpha$  fluorescence line in the spectra of Seyfert 1 galaxy MCG-6-30-15 \citep{b32}. Similar studies had been performed on samples of local AGN (see e.g. \citet{b23,b33,b24,b19}, and some distant quasars \citep{b4} leading to the understanding of the properties of the Fe~K$\alpha$  fluorescence line emission.	

The 6.7 and 7.0 keV emission lines, in contrast, are  produced by either photoionization or collisional excitation in hot plasma, depending upon the source. In AGN and black hole binaries these lines are largely produced by photoionization due to intense X-ray luminosity from the compact objects, while in neutron stars and white dwarf binary systems collisional processes from the accretion process can dominate.
 
Hard X-ray emitting symbiotic stars (hereafter hSSs) such as CH~Cyg, T~CrB, RT~Cru and SS73~17 observed with Suzaku satellite were found to emit strong Fe~K$\alpha$ fluorescence emission line at 6.4~keV  with an average equivalent width of 180 eV and the 6.7 and 7.0 keV lines with equivalent width of 88 and 74 eV respectively \citep{b10}. The question this work will address is what are the origin of these emission lines?   In order to answer this question we carried out spectral re-analysis of the Suzaku observations on these systems and our result supports the earlier results that the possible origin of the hard X-rays emitted from the hSSs is from the boundary layers between the accretion disk and accreting white dwarf \citep{b16,b8}.

\section{Data Selection}
The hSSs were selected based on the fact that they have been observed with Suzaku and were confirmed to have strong Fe~K$\alpha$\ emission lines with hard-tails above 20~keV. Four hard X-ray emitting symbiotic stars, SS73~17, RT~Cru, T~CrB, and CH~Cyg  (e.g., \citet{b13,b8}) were selected and used for our study (Table~1).

\section{Data Analysis and Results}
Analysis of our data were done using version~2 of the standard Suzaku pipeline products, and 
the HEASoft\footnote{%
  See http://heasarc.gsfc.nasa.gov/lheasoft/ for details. 
}
version~6.10. 
In majority of the sources we used $250\arcsec$~radius to extract all events for the XIS 
detector for the production of the source spectra but in some cases where the $250\arcsec$~radius overlaps with the calibration sources at the corners, we adjusted the radius accordingly. 
The XIS background spectra were extracted with $250\arcsec$~radius with no apparent sources 
and were offset from both the source and corner calibrations. 
The radius was also adjusted accordingly in some cases where it overlaps with the calibration 
sources at the corners. 
The RMF and ARF files for the spectral analysis were generated for the XIS detector using the 
FTOOLS \texttt{xisrmfgen} and \texttt{xissmarfgen} respectively. 
Suzaku XIS 0, 2, and 3 have front-illuminated (FI) chips with similar features, so we merged the 
spectra of XIS 0 and 3, which we hereafter refer to as XIS FI (XIS~2 has been out of service 
since November 9, 2006 due to an anomaly). 
XIS~1 is back-illuminated (BI), and we hereafter refer it to as XIS BI. 

The HXD PIN detector analysis was done by first downloading the non X-ray tuned PIN background (NXB)
files and response matrix appropriate for each observation as generated by Suzaku team. 
We used the \texttt{mgtime} FTOOLs to merge the good time intervals to get a common value for 
the PIN background and source event files. 
The source and background spectra extraction were done for each observation using the 
\texttt{xselect} filter time file routine. 
We corrected for the dead time of the observed spectra using the \texttt{hxddtcor} in the 
Suzaku FTOOLS. 
According to the standard analysis procedure, the exposure time for all observations for the derived 
background spectra were increased by a factor of 10 to take care of the event rate in the PIN 
background event file which is 10 times higher than the true background in order to suppress Poisson errors. We used the appropriate Cosmic X-ray Background (CXB) flat response files as provided by the Suzaku team to create CXB spectra. We then merged the NXB and CXB spectra using \texttt{mathpha}.

Spectral analysis of all observations were performed using XSPEC version~12.7.0. 
We modeled the spectrum using an absorbed bremsstrahlung model with three Gaussian lines for 
the three Fe~K$\alpha$ emission lines. 
Since we were primarily interested in the ion lines our fitting covers 3--10~keV for the XIS BI, 
3--12~keV for the XIS FI and 15--40~keV for the HXD PIN. 
We ignored the energy range below 3~keV in the XIS FI and BI detector to avoid intrinsic absorption 
which is known to affect data at this energy range and energies above 10~keV were ignored for 
XIS BI because the instrument background is higher compared to the XIS FI detectors. 

The three  lines, from (near-) neutral (6.4~keV), He-like (6.7~keV), and H-like (7.0~keV) iron ions were clearly resolved in all the sources. Spectra of all the sources were presented in Figure~1 and spectral parameters were shown in Table~3. 

We detected strong 6.4~keV iron line emission in all the hSSs spectrum with an average equivalent  width (EW) of 221~eV and strong 6.7 and 7.0 keV lines with average equivalent width of 92 and 75~eV, respectively.

The values of the reduced  $\chi^2$\  for the absorbed bremsstrahlung, power law plus XSPEC reflect model and absorber bremsstrahlung plus XSPEC reflect models were presented in Table 2 . Our modeling of the spectrum of our sources with power law plus XSPEC reflect model with three Gaussian lines for the iron lines did not produce a statistically acceptable fit.  This is not surprising since our knowledge of hard X-ray emitting symbiotic stars shows that we always require a thermal model to obtain a statistically acceptable fit at the lower energy band (0.1--10 keV) ( see \citet{b7,b16,b30,b13,b8}). However, at higher energy bands (15--50 ~ keV) a non-thermal spectrum can be detected in some sources. As an example the INTEGRAL observations of RT Cru were well fit with non-thermal power law emission with photon index G = 2.7 \citep{b3}. This implies that non-thermal radiation mechanism could possibly be present in our sources, but is not apparent except at the higher energy band. We confirmed the presence of reflection component in one of our sources SS73~17, from modeling our sources with an absorbed bremsstrahlung plus XSPEC reflect model (see Table 2 for the reduced $\chi^2$\ values) 
\section{Discussion and Conclusion}

\subsection{On the origin of hard X-rays and the 6.4 keV line from hSSs}

Symbiotic stars are generally known to emit soft X-rays, which are believed to be resulting from materials burning quasi-steadily on the white dwarf WD surface \citep{b12,b25}. A small class of symbiotic stars emit hard X-rays, SS73 17, RT Cru, CH Cyg, T CrB, and MWC 560. The origin of hard X-rays from hSSs had been addressed by some authors \citep{b16,b13,b8} which points to the boundary layer between the accreting white dwarf and the accretion disk.  This seems most likely since no other area or mechanism in the system can produce such high energy X-rays. Other possible sources, such as magnetic reconnection in the base of jet or an expanding shock front are not likely to be present in a white dwarf / red giant system. The detection of strong iron lines  and most particularly the thermal bremsstrahlung continuum with an average temperature of 20~keV in all the sources strongly supports the thermal origin of the hard X-rays, and in particular the collisional origin of the He-like and H-like 6.7 and 7.1 keV lines. Hence, these hard X-rays are most likely released during the accretion process as material from the red giant companion falls down towards the white dwarf and stops at the boundary layer. We note as further evidence for this picture that we observed high absorption of the hard X-rays with both full covering and partial covering in all the sources (see Table~3), consistent with a large accretion column covering the hot interaction region.

The 6.4~keV fluorescence line emission is usually caused by irradiation of (near-) neutral material (in this case, iron) by a hard X-ray source, ejecting one of the 2 K-shell ( n = 1) electrons of an Fe atom (or ion). While electron collisions could also cause such an ejection, and in hSSs a $\sim 20$\,keV collisional plasma is often seen, this is unlikely to be origin as a collisional plasma would rapidly ionize the iron.  The more likely scenario is that there is some separation between the energetic X-ray photon-emitting plasma and the (near-) neutral Fe atoms.  

One K-shell electron is ejected following photoelectric absorption of a hard X-ray, the resulting excited state can decay in one of two ways. An L-shell (n = 2) electron may drop into the K-shell releasing $\sim 6.4$\,keV of energy either as a fluorescence emission line photon or an Auger electron.  An Auger electron occurs when the energy produced by the n = 2 $\rightarrow$ n = 1 transition is internally absorbed by another electron which is ejected from the ion.). The $\sim 6.4$\,keV line is actually a doublet, with slightly different energies depending on spin-orbit interaction energy between the electron spin and the orbital momentum of the $2p$\ orbital. These two line components of the K$\alpha$ line are at 6.404 and 6.391 keV, though they are  not separately distinguished at CCD resolution. The hSSs are binary systems, with a white dwarf accreting matter from an accretion disk, though some hSSs are believed to have a white dwarf accreting matter from a cocoon of gas surrounding it and coming from the red giant companion (see e.g. \citet{b31,b34}). 

\subsection{Case I: The Accretion Disk / Reflection Origin}

The result of modeling our sources with addition of the XSPEC reflect model to an absorbed bremsstrahlung shows that there is none or low reflection of hard X-rays in CH~Cyg, T~CrB and RT~Cru, while for SS73~17 there is significant indication that hard X-rays from the boundary layers between the accretion disk and the compact objects ( see e.g., \citet{b16,b8}) irradiate the cold gas in at least half of the surface of the accretion disk leading to the emission of the Fe~K$\alpha$ fluorescence line (see table 2). Interestingly two of our sources CH~Cyg and T~CrB has absorption column densities of the order of $N_{\mathrm{H}}$ $\sim$ $10^{23}$~cm$^{-2}$ in both full--covering and partial--covering matter, while for RT-Cru the absorption column densities is of the order of $N_{\mathrm{H}}$ $\sim$ $10^{22}$~cm$^{-2}$ for the full--covering matter and  it is in $N_{\mathrm{H}}$ $\sim$ $10^{23}$~cm$^{-2}$ for the partial--covering matter (see table 3). The mechanism for the creation of the Fe~K$\alpha$ fluorescence line in such system will be discussed in the next section. In the case of hSSs that accrete matter directly from the wind of  the red giant, hard X-rays released from the surface of the white dwarf during the accretion of matter, irradiates matter in the thick cold gas surrounding the white dwarf thereby emitting the 6.4 keV line. 

Generally, for some compact objects that accrete matter through the accretion disk process such as the white dwarfs, neutron stars, and black holes, there are always the emission of Fe~K$\alpha$ fluorescence line as well as Compton reflection, both signifying the presence of cold matter. Fe~K$\alpha$\ fluorescence lines were detected in Seyfert galaxies with typical EW of $\sim$ 200--300 eV using ASCA \citep{b23,b24}. Compton reflection was earlier discovered also in Seyfert galaxies by Ginga \citep{b21,b27,b29} and the average solid angle $\Omega$ subtended by the reflecting medium was $\Omega$/2$\pi$ = 0.5 \citep{b23}.  The range of possible EW depends upon the geometry of the source.  If the ionizing source itself is not hidden from view, then the largest possible EW is $\sim 200$\,eV \citep{b11} for a face-on slab of material with typical iron abundances.  This value can be exceeded if the source is hidden from view by an optically thick absorber and only the reflected light observed, but such a situation is more likely to occur in an AGN with a large surrounding torus than in a hSS with only the accreted matter from red giant wind to provide absorption.  We note that three of our four sources all have 6.4 keV EW's under 200 eV (see Table 3), although this is only corroborating evidence for the reflection / accretion disk geometry picture. 

Compton reflection plays a vital role in the X-ray spectrum of many compact objects especially in active galactic nuclei (AGN) and black hole candidates (BHC) sources. When the primary X-rays hits Thomson-thick matter (e.g. the accretion disk), Compton--scattered X-rays enter into our line of sight. Elements like iron, which are in low state of ionization, can absorb the lower energy X-rays and so flatten the reflection continuum leading to the production of the fluorescent line emission \citep{b18,b11} In most AGN such as Seyfert 1 galaxies, the basic spectral model is that of a hard X-ray emitting continuum source illuminating an accretion disk, which leads to the observed reflected hard X-ray emission together with soft X-rays from thermal emission disk. However in some AGN sources (e.g. \citet{b15,b6,b27}), with large iron line equivalent width of the order of 300--400 ~eV, the basic model does not apply. This may well explain why our hard X-ray emitting symbiotic stars with relatively large iron line equivalent width of the order of 117--580~eV could not be modeled using the power law plus reflection component model at the 0.1--10~keV energy band. However, hSSs could be modeled with power law plus reflection component at higher energy band 15--50 ~keV (see \citet{b3}). In our further search for a reflection component in our sources, we successfully modeled three out of our four sources with an absorbed bremsstrahlung plus reflection component (see Table 2 for the reduced $\chi^2$\ values). This strongly suggests the presence of reflection of Compton scattered hard X-rays from the accretion disks leading to the formation of the fluorescence iron line.

\subsection{Case II: The Cold Absorber Option}

Hard X-rays traveling through absorbing material will interact with the iron, just as in the reflection case, to create Fe K$\alpha$\,fluorescence lines.  In this case, the fluorescence iron line intensity will increase with the thickness $N_{\mathrm{H}}$ of the ambient matter.  Of course, when $N_{\mathrm{H}}$ becomes larger than (3--5) $\times$ $10^{23}$~cm$^{-2}$, the line intensity starts to decrease due to self-absorption of photons by the matter \citep{b20}. Iron K$\alpha$ lines which were observed in few AGN that showed absorption column densities of $N_{\mathrm{H}}$ $\sim$ $10^{23}$~cm$^{-2}$\ such as Centaurus A \citep{b22} and NGC 4151 \citep{b26} has been linked to the fluorescence from the same matter in the absorption column \citep{b28}.

The large absorption column density of the order of $10^{23}$~cm$^{-2}$ in the hSSs guarantees that some of the iron fluorescent line will arise from interactions between the hard X-rays with the thick column density of cold matter in our line of sight (see e.g. \citet{b21}).  This picture is enhanced by a review of Table 3, which shows that the largest Fe K$\alpha$\ EW of $580^{+424}_{-65}$\,eV, seen in CH Cyg, is also associated with the largest actual column density, using either the observed full-covering value or the sum of the total and the weighted partial covering fraction as a proxy for the N$_{\rm H}$.  Piro (1993) found that the EW generated by passage through optically-thin absorbing material is $\sim100 {\rm eV\ } \times A_{Fe} \times {\rm N}^{23}_{\rm H} \times f_{\Omega}$, where $A_{Fe}$ is the Fe abundance to solar, ${\rm N}^{23}_{\rm H}$\ is the column density in units of $10^{23}$\,cm$^{-2}$\ and $f_{\Omega}$\ is the total covering fraction of the absorber. In the case of CH Cyg, this would create an EW of either 220 eV (using only the full fraction) or 1120 eV using the weighted value.

In contrast to the CH Cyg situation, however, the other three sources show a much lower correlation between column density and the Fe K$\alpha$\ EW.  T CrB has the lowest EW of the three sources but a significantly larger column density, measured either with with full-covering N$_{\rm H}$\ or via the weighted method, than either RT Cru or SS73 17.  Of course, these two methods are merely approximate; a full calculation of the expected EW would require complete modeling of the geometry of the source and absorption system, including the arrangement of the ``partial'' covering. At the least, however, this discrepancy suggests that absorption is only one part of the observed EW in these systems.

\subsection{The 6.7 and 7.0 keV Emission Lines}

The He-like and H-like iron emission lines can be produced both by photoionization (and excitation) and collisional ionization/excitation in hot plasmas. It is a matter of which one dominates in any given source, which may not be simple to deduce.  In the case of SS73 17, \citet{b8} used high-resolution Chandra HETG to show that the G ratio in the Fe XXV lines, albeit heavily blended, was much more consistent with that of a collisional plasma than a photo ionized one.  In addition, they showed that the H-like to He-like line ratio was consistent with the best-fit bremsstrahlung temperature of the system.   Unfortunately, the accuracy of our He-like and H-like EW measurements is not adequate to constrain their ratios sufficiently to merit comparison with the measured bremsstrahlung temperature.  In addition, the Suzaku CCDs do not have sufficient resolution to measure the G ratio from the Fe XXV line itself.  However, we believe it is likely that the emission lines from the other three sources are dominated by thermal emission as well, although this could and should be tested with future Chandra HETG or Astro-H observations.

The author wishes to acknowledge the Suzaku team for providing data and some relevant files used in the analysis of this work. He is very grateful to the Japan Society for the Promotion of Science (JSPS) for financial support under the JSPS Invitation Fellowship (Long Term) and ISAS/JAXA, Sagamihara Campus for hosting him. The author wishes also to thank Ken Ebisawa, his host researcher and Kei Saito for their useful discussions during the time of this research, and to Randall Smith who assisted significantly in many areas. The author also acknowledges the extremely helpful comments and suggestions made by the anonymous referee. This research made use of data obtained from Data ARchives and Transmission System (DARTS), provided by Center for Science-satellite Operation and Data Archives (C-SODA) at ISAS/JAXA. 

%


%

\begin{table*}
  \caption{The symbiotic stars used in the work}
 \begin{tabular}{@{}lccc}
  \hline
  Source Name & ObsID & Obs. start (UT) & Exp. \\
   && Date / Time & (ks) \\
  \hline
  CH~Cyg & 
    400016020 & 2006-05-28 / 07:28 & \phantom{0}35.2 \\
  T~CrB & 
    401043010 & 2006-09-06 / 22:44 & \phantom{0}46.3 \\
  RT~Cru & 
    402040010 & 2007-07-02 / 12:38 & \phantom{0}50.9 \\
  SS73~17 & 
    403043010 & 2008-11-05 / 16:30 & \phantom{0}24.9 \\
 \hline
\end{tabular}
\end{table*}

\begin{table*}
  \caption{The reduced  $\chi^2$\  for the absorbed bremsstrahlung, power law plus XSPEC reflect model and an absorbed bremsstrahlung plus XSPEC reflect model, F--test probability and reflection normalization}
 \begin{tabular}{@{}lccccc}
  \hline
  Source Name & BM & PRM & BRM & FP & RN  \\
    \hline
  CH~Cyg & 
    1.40 & \phantom{0}5.56 & \phantom{0}1.40 & \phantom{0}0.29 & \phantom{0}0.00 \\
  T~CrB & 
    1.19 & \phantom{0}2.53 & \phantom{0}1.18 & \phantom{0}0.05 & \phantom{0}0.06 \\
  RT~Cru & 
    1.14 & \phantom{0}1.79 & \phantom{0} 1.12 & \phantom{0}0.002 & \phantom{0}0.17 \\
  SS73~17 & 
    1.20 & \phantom{0}2.57 & \phantom{0} 1.02 & \phantom{0}$3.61 \times 10^{-8}$ & \phantom{0}0.46 \\
 \hline
\end{tabular}
 \medskip 
    Parameters are 
    the absorbed bremsstrahlung model (BM), power law plus XSPEC reflect model (PRM), an absorbed bremsstrahlung plus XSPEC reflect model (BRM), F--test probability of null result for the BRM (FP) and reflection normalization for the BRM (RN)
    \end{table*}

\begin{table*}
\centering
  \caption{The Fe lines fit parameters.}
 \begin{tabular}{@{}lcccccccccc}
  \hline
  Source Name &
    $N_{\mathrm{H}}^{\mathrm{f}}$ & $N_{\mathrm{H}}^{\mathrm{p}}$ & $C$ &
    $kT$ &
    $E_{\mathrm{6.4}}$ & EW$_{\mathrm{6.4}}$ &
    $E_{\mathrm{6.7}}$ & EW$_{\mathrm{6.7}}$ &
    $E_{\mathrm{7.0}}$ & EW$_{\mathrm{7.0}}$ \\
   \hline
   CH~Cyg & 
    $22.0\pm7.0$ & $99_{-37}^{+26}$ & $0.91\pm0.04$ & 
    $6_{-2}^{+3}$ & $6.41\pm0.02$ & $580_{-65}^{+424}$ & 
    $6.59\pm0.05$ & $111_{-35}^{+95}$& $6.81_{-0.07}^{+0.12}$ 
    & \phantom{0}$66_{-38}^{+87}$ \\
  T~CrB & 
    $17.7\pm2.2$ & $37\pm6$ & $0.71\pm0.06$ & 
    $19_{-2}^{+3}$ & $6.43\pm0.01$ & $117_{-46}^{+48}$ & 
    $6.72\pm0.02$ & $85_{-12}^{+40}$ & $7.01\pm0.02$ 
    & \phantom{0}$104_{-41}^{+38}$ \\
  RT~Cru & 
    $3.3\pm0.4$ & $58\pm9$ & $0.46\pm0.05$ & 
    $29_{-5}^{+9}$ &  $6.38\pm0.01$ & $174_{-30}^{+38}$ & 
    $6.64\pm0.01$ & $52_{-15}^{+34}$ & $6.96\pm0.01$ 
    & \phantom{0}$51_{-25}^{+17}$ \\
  SS73~17 & 
    $9.2\pm0.5$ & $34\pm5$ & $0.65\pm0.05$ & 
    $38\pm6$ & $6.38\pm0.01$ & $185_{-63}^{+120}$ & 
    $6.67\pm0.01$ & $158_{-41}^{+136}$ & $6.95\pm0.01$ 
    & \phantom{0}$93_{-40}^{+116}$ \\
    \hline 
    \end{tabular}
     \medskip 
    Parameters are 
    the hydrogen column density of the full-covering and the partial-covering matter in units 
    of $10^{22}$~cm$^{-2}$ ($N_{\mathrm{H}}^{\mathrm{f}}$ and $N_{\mathrm{H}}^{\mathrm{p}}$), 
    the covering fraction of the partial-covering matter ($C$), 
    the continuum temperature in keV ($kT$),
    the center energies of 6.4, 6.7, and 7.0~keV lines in keV ($E_{\mathrm{6.4}}$, 
    $E_{\mathrm{6.7}}$, and $E_{\mathrm{7.0}}$),
    the equivalent widths in eV (EW$_{\mathrm{6.4}}$, EW$_{\mathrm{6.7}}$, and EW$_{\mathrm{7.0}}$). 
 \end{table*}
\newpage
%


%

%
\begin{figure*}
 \begin{minipage}{70truemm}
  \includegraphics[width=84mm]{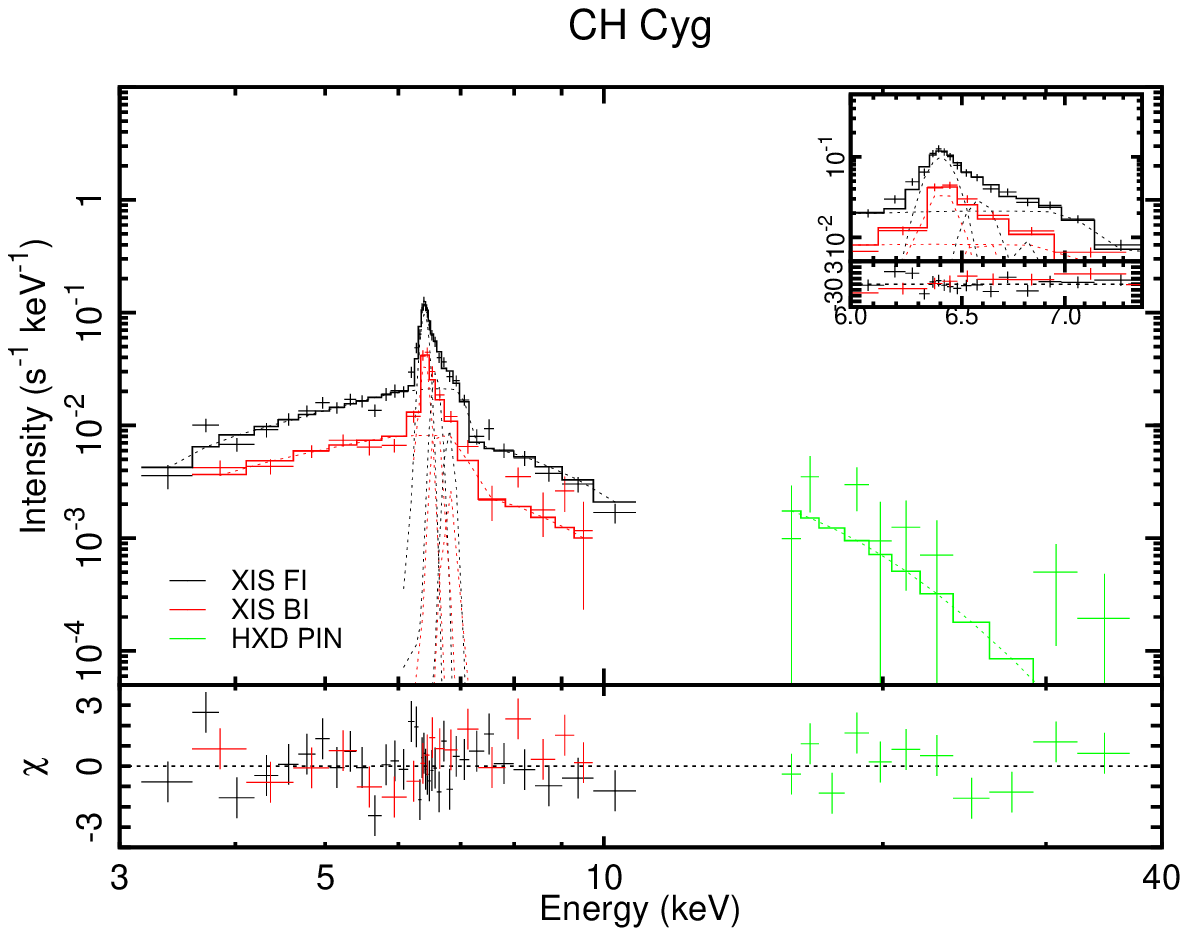}
  \end{minipage}
 \hspace{20truemm}
 \begin{minipage}{70truemm}
  \includegraphics[width=84mm]{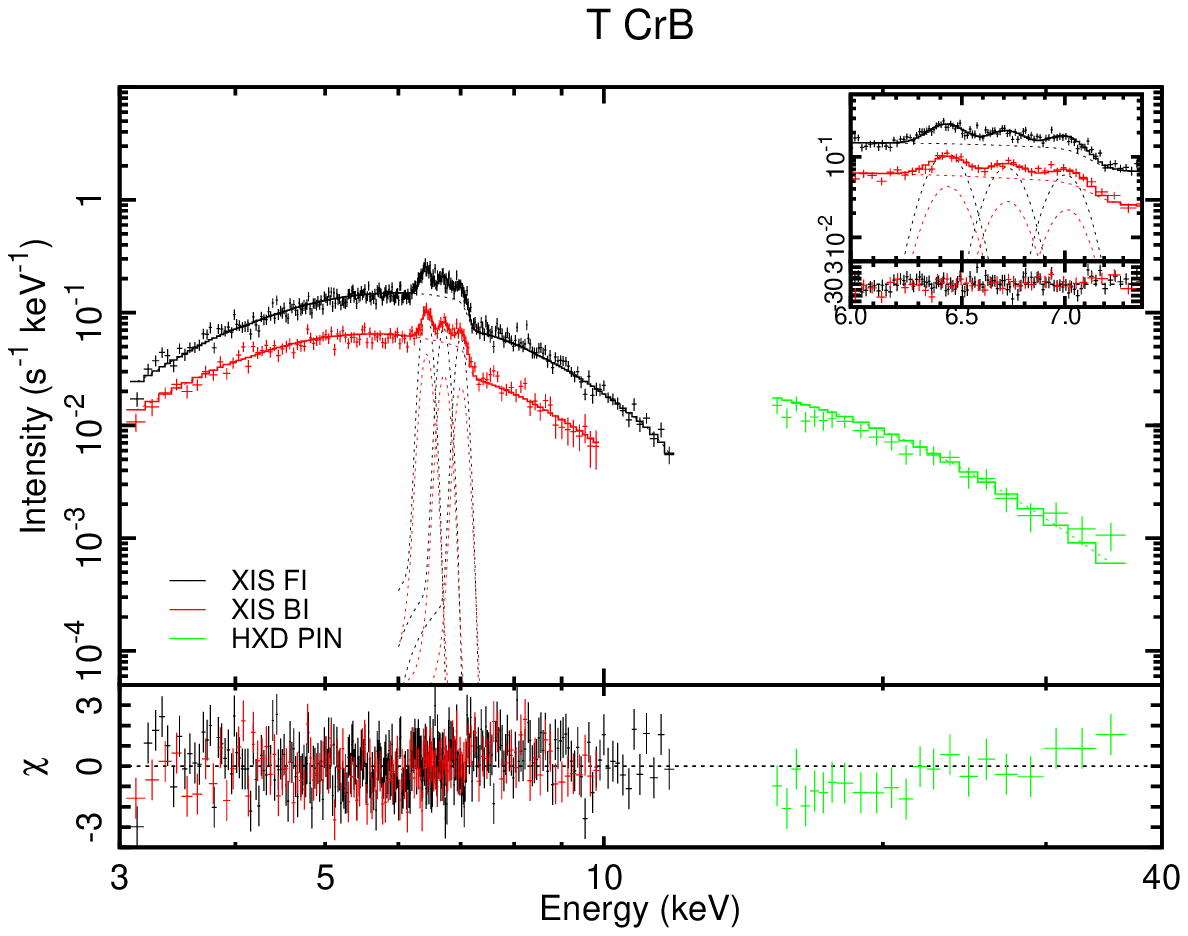}
 \end{minipage}
 \begin{minipage}{70truemm}
 \vspace{10truemm}
  \includegraphics[width=84mm]{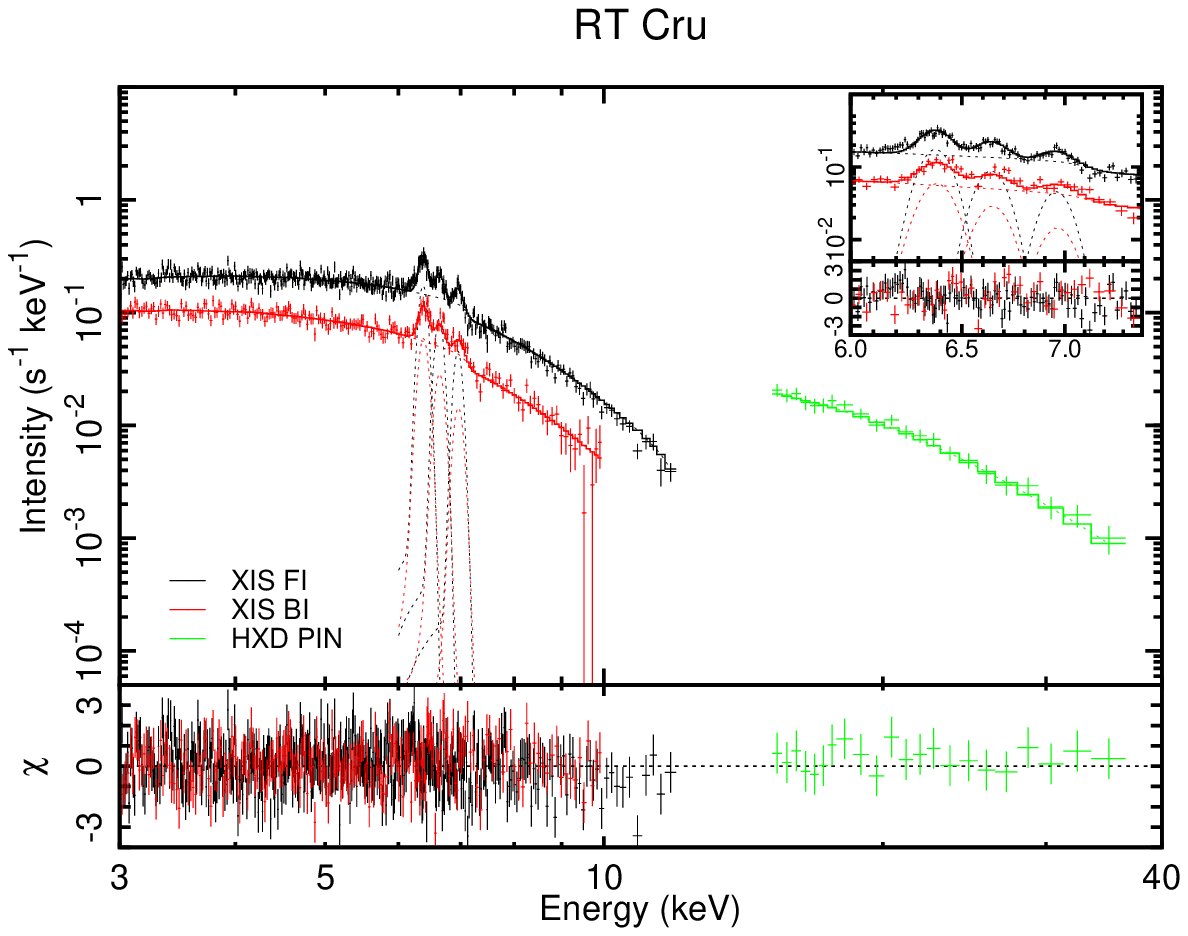}
 \end{minipage}
 \hspace{20truemm}
 \begin{minipage}{70truemm}
 \vspace{10truemm}
  \includegraphics[width=84mm]{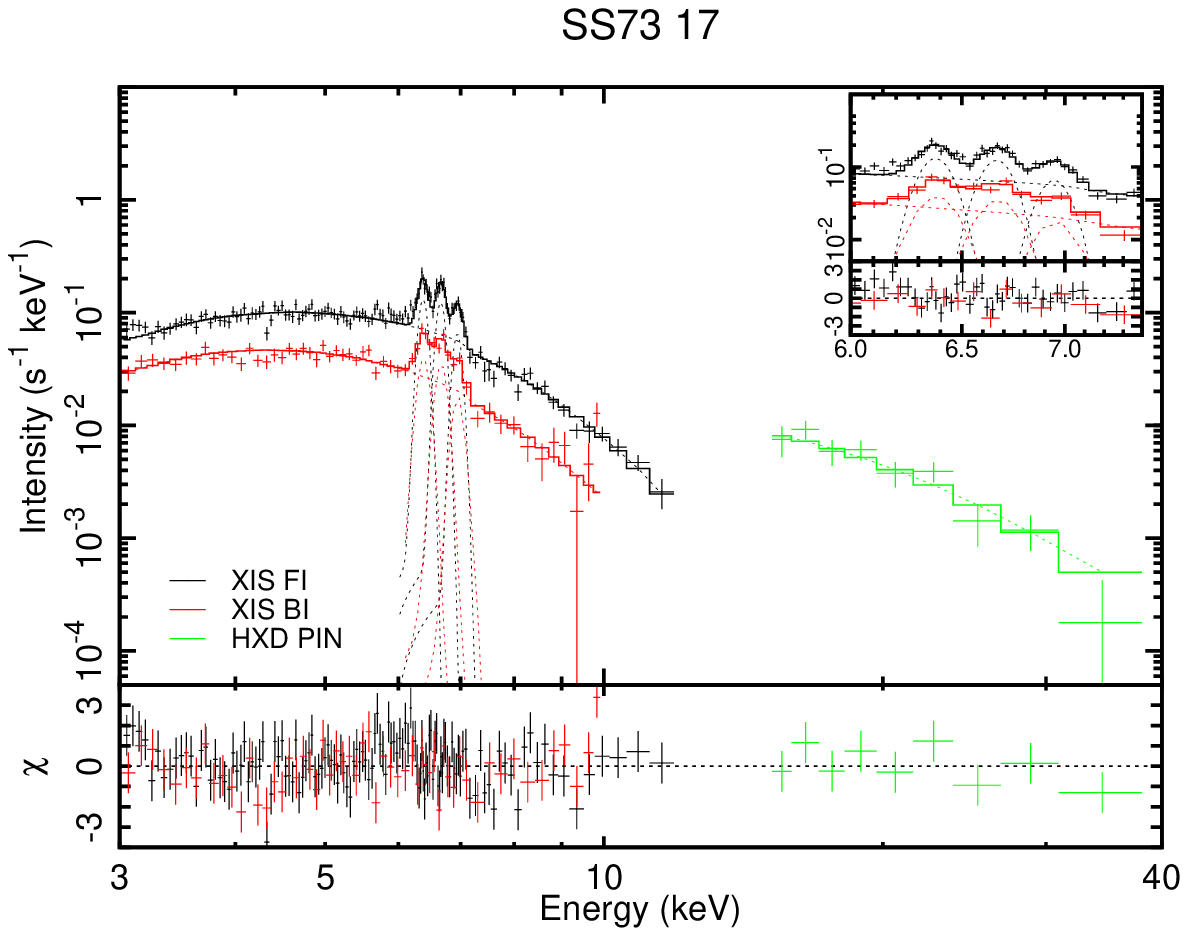}
 \end{minipage}
  \caption{Spectra of the symbiotic stars. 
           In the upper panel, the data and the best-fit model are shown by crosses and solid 
           lines, respectively. 
           Each spectral component is represented by dotted lines. 
           In the lower panel, the ratio of the data to the best-fit model is shown by crosses. 
           The inset in the upper panel is an enlarged view for the Fe~K$\alpha$ complex lines.}
  \label{f1}
\end{figure*}
 \end{document}